\documentstyle[aps,twocolumn,epsf]{revtex}
\begin{document}
\draft
\def \beq{\begin{equation}}
\def \eeq{\end{equation}}
\def \beqarr{\begin{eqnarray}}
\def \eeqarr{\end{eqnarray}}

\title{Quantum Hall Skyrmions with Higher Topological Charge}

\author{D. Lillieh\"{o}\"{o}k$^1$, K. Lejnell$^1$, A. Karlhede$^1$, and 
S. L. Sondhi$^2$.}

\address{
$^1$Department of Physics,
Stockholm University,
Box 6730, S-11385 Stockholm,
Sweden
}

\address{
$^2$Department of Physics,
Princeton University,
Princeton, NJ 08544, USA
}

\date{\today}
\maketitle
\begin{abstract}
We have investigated quantum Hall skyrmions at filling factor $\nu=1$
carrying more than one unit of topological, and hence electric, charge. 
Using a combination
of analytic and numerical methods we find the counterintuitive result
that when the Zeeman energy is tuned to values much smaller than the 
interaction energy ($g \mu_B B/(e^2/\epsilon\ell) < 9 \times 10^{-5}$), 
the creation 
energy of a charge two skyrmion becomes less than twice the creation 
energy of a charge one skyrmion, i.e. skyrmions bind in pairs. The doubly 
charged skyrmions are stable to further accretion of charge and exhibit a 
10 \% larger spin per unit charge than charge one skyrmions which would, in 
principle, signal this pairing.
\end{abstract}
\pacs{}

Ferromagnetic quantum Hall (QH) systems have quasiparticle excitations,
``skyrmions'', that involve a texturing of the spins \cite{sondhi1}. There is 
now strong experimental evidence that the lowest energy charged excitations at
filling factor $\nu=1$ are skyrmions \cite{barrett,schmeller,goldberg,eaves}. 

A smooth texturing of the spins can be described by an effective non-linear
sigma model (see (\ref{effective}) below) where the spin is represented 
by a unit vector ${\bf n}({\bf r})$. Skyrmions are topological excitations 
characterized by an integer topological charge $Z=\int d{\bf r} q({\bf r})$, 
where $ q={\bf n}\cdot (\partial _x {\bf n} \times \partial _y {\bf n})/4\pi$ is the 
topological (Pontryagin) density; $Z$ is the winding number of the mapping 
${\bf n}({\bf r})$ from the compactified plane ($S_2$) to the target space of
the sigma model ($S_2$). A characteristic feature of QH skyrmions is that the 
topological density is proportional to the deviation of the electron density, 
$\rho$, from its  uniform value, $\bar \rho$: $q= \nu(\rho-\bar \rho)$. Thus,
at $\nu=1$, skyrmions with topological charge $Z$ carry electric charge $-Z|e|$.

Up till now only skyrmions with charge one have been considered for the natural
reason that one would expect Coulombically repelling skyrmions to disfavor
charge aggregation. In this note we study skyrmions at $\nu=1$ with higher charge
and find that this expectation is violated for extremely small values of the Zeeman
energy where the skyrmions are large objects. More precisely, we find that a charge 
two skyrmion has lower energy than a pair of charge one skyrmions for $\tilde g < 
\tilde g_c =9 \times 10^{-5}$, where $\tilde g \equiv g \mu_B B/(e^2/\epsilon \ell)$ 
is the
dimensionless Zeeman energy. This implies that the interaction between two skyrmions,
which must be repulsive at long distances on account of their electric charges,
has an attractive core in a region where their identities are no longer distinct.
We also find that skyrmions with charge three and higher are energetically disfavored
at all values of $\tilde g$.

We begin with variational estimates of the skyrmion energies for various 
topological charges at small $\tilde g$ that were the motivation for
this work. We then study the skyrmions numerically, both in the effective 
sigma model by integrating the equations of motion \cite{kennet}, and more
microscopically using a Hartree-Fock scheme \cite{fertig}. These establish
the result quoted above and determines $\tilde g_c$. Finally, we discuss
the prospects for observing this effect in experiments.

\noindent
{\bf Effective Sigma Model:} The long wavelength physics of the spin degrees 
of freedom in ferromagnetic QH states is described by the effective Lagrangian
density \cite{sondhi1}
\begin{eqnarray}
\label{effective}
{\cal L}_{eff} = &\frac 1 2& {\bar \rho} {\bf {\cal A}}({\bf n}) \cdot 
\partial_t {\bf n}
- \frac{1}{2}\rho^s (\nabla {\bf n})^2
\nonumber \\
+ &\frac 1 2& g \overline{\rho} \mu_B {\bf n} \cdot {\bf B}
-\frac{e^2}{2 \epsilon} \int d^2r' \frac{q({\bf r}) 
q({\bf r}')} {|{\bf r}-{\bf r}'|} \ \ \ .
\end{eqnarray}
Here ${\bf {\cal A}}$ is the vector potential of a unit magnetic monopole, 
$\rho^s$ is the spin stiffness ($\rho^s = e^2/(16\sqrt{2 \pi}
\epsilon \ell)$ 
for $\nu=1$),
$\epsilon$ is the dielectric constant and $\ell$ the magnetic length.

The first two terms in ${\cal L}_{eff}$ are the leading terms in the effective
action for any ferromagnet and on account of the scale invariance of the
gradient term lead, {\em prima facie}, to charge $Z$ skyrmions of arbitrary 
size $\lambda$ and energy $E_Z=4\pi |Z| \rho^s$ \cite{rajaraman}. 

The Zeeman and Coulomb terms in  ${\cal L}_{eff}$ break the scale invariance
and their competition sets the  size and energy of the skyrmions and
modifies the detailed form of their profiles; these now depend on the dimensionless 
ratio $\tilde g = (g\mu_B B)/(e^2/ \epsilon \ell ^2)$ of the Zeeman energy to the 
Coulomb energy. 
A rotationally symmetric ansatz for a skyrmion with topological charge $Z$ is 
\begin{eqnarray}
  n_x &=&  \sqrt{1 - f^2(r)} \, \cos(Z\theta) \ , \nonumber \\
  n_y &=&  \sqrt{1 - f^2(r)} \, \sin(Z\theta) \ , \ \ \
  n_z = f(r)  \ \ \ , 
\label{ansatz}
\end{eqnarray}
where $f$ obeys the boundary conditions $f(0)=-1, \, f(\infty)=1$. The 
topological density is $q=(Z/4\pi r)df/dr$. The skyrmions in the scale 
invariant sigma model are given by $f(r)=[(r/\lambda)^{2Z}-4]/[(r/\lambda)^{2Z}+4]$. 
For $Z>1$, their topological density has a hollow core. 

Substituting (\ref{ansatz}) in the action (\ref{effective}) and minimizing
leads to a non-linear, non-local integro-differential equation for $f$.
Below, we discuss results obtained by numerically integrating this
equation using a relaxational technique \cite {kennet}. But first, it
is instructive to consider approximate solutions at small $\tilde g$;
the details are discussed elsewhere \cite{kennet}. For $Z >1$ we take
the solution to be of the form of the scale invariant sigma model
solution with an optimized scale parameter $\lambda$. This yields,
\beqarr
\label{charge234}
\lambda &=& 0.780 \ell \tilde g^{-1/3}\ {\rm and}\ E= {e^2\over\epsilon \ell} \big[
2 \sqrt{\pi \over 32} + 2.86 \tilde g^{1/3} \big]\ \ (Z=2) \nonumber \\
\lambda &=& 1.29 \ell \tilde g^{-1/3}\ {\rm and}\ E= {e^2\over\epsilon \ell} \big[
3 \sqrt{\pi \over 32} + 4.82 \tilde g^{1/3} \big]\ \ (Z=3) \nonumber \\
\lambda &=& 1.75 \ell \tilde g^{-1/3}\ {\rm and}\ E= {e^2\over\epsilon \ell} \big[
4 \sqrt{\pi \over 32} + 7.21 \tilde g^{1/3} \big]\ \ (Z=4)
\eeqarr
for the lowest three values of $Z$. Note that the energy per unit charge is
the same for the $\tilde g$ independent piece as is appropriate for
solutions of the scale invariant problem \cite{rajaraman} while for the
$\tilde g$ dependent piece it increases monotonically with $Z$. Hence, there
is no binding between this set of skyrmions.

However, this procedure runs into trouble with the $Z=1$ skyrmion where the
scale invariant solution yields a Zeeman energy that diverges logarithmically
with system size for any $\lambda$. This can be fixed by matching the scale
invariant solution onto the exact asymptotic solution in the outer region which
decays exponentially with a length $\propto \ell \tilde g^{-1/2}$. As a 
consequence we get logarthmically modified expressions \cite{sondhi1,meaculpa},
\beqarr
\lambda &=& 0.558 \ell (\tilde g |\ln \tilde g|)^{-1/3}\ {\rm and} \nonumber \\ 
E &=& {e^2\over \epsilon \ell} \big[ \sqrt{\pi \over 32} + 
0.622 (\tilde g |\ln \tilde g|)^{1/3} \big]\ \  (Z=1) \ \ ,
\label{charge1}
\eeqarr
for the core scale parameter and energy. It is evident that the presence of
the logarithm by itself implies that charge one skyrmions will pair bind for 
small enough $\tilde g$  and that the binding energy will 
vanish as $\tilde g$ vanishes. Using the above expressions one finds a critical
value of $\tilde g_c \approx 5.3 \times 10^{-6}$ for this to happen. That such a critical
value has to exist can be inferred by a separate microscopic
computation of the pair-binding energy at large $\tilde g$, where the skyrmion 
size is $O(\ell)$ and they reduce to polarized quasiparticles, which shows the
abscence of binding. However, the quantitative issue is rather delicate.
Even at small $\tilde g$ the result (\ref{charge1}) is expected to have 
logarithmically subdominant corrections and is known numerically to be not 
terribly accurate in estimating the {\em difference} $E(\tilde g) - E(0)$, 
even for $\tilde g$ as small as $10^{-6}$ \cite{kennet}. Also, the asymptotic
validity of our small $\tilde g$ expressions is not on completely rigorous
footing at present. So it follows that while the pair binding should be present 
only at small $\tilde g$, a reliable estimate of the critical $\tilde g_c$ 
requires the more accurate computations of the skyrmion energies reported below.

\noindent
{\bf Hartree-Fock:} We complement the effective action calculations by
the Hartree-Fock method introduced by Fertig {\it et. al.} \cite{fertig}.  
For skyrmions with topological charge $Z$ the Hartree-Fock wave function 
is fixed by requiring it to be in the lowest Landau level and by the 
symmetries of the classical solution (\ref{ansatz}) to be
\begin{eqnarray}
\label{hfstate}
|\Psi \rangle &=& \prod _{m=0}^{\infty} (u_m c^{\dagger}_{m\uparrow} + v_m 
c^{\dagger}_{m+Z \downarrow})\prod _{p=0}^{Z-1}c^{\dagger}_{p \downarrow}
 |0 \rangle \ \ \ {\rm if} \ \ \  Z > 0 
\nonumber  \\
|\Psi \rangle &=& \prod _{m=-Z}^{\infty} (u_m c^{\dagger}_{m\uparrow} + v_m 
c^{\dagger}_{m+Z \downarrow})
 |0 \rangle \ \ {\rm if} \ \ \  Z < 0 \ \ \ \ ,
\end{eqnarray}
where $|u_m|^2+|v_m|^2=1$,  $c_{m\sigma}|0\rangle=0$ and 
$ c^{\dagger}_{m\sigma}$  creates  electrons with spin $\sigma$
in the  angular momentum $m$-state in the lowest Landau 
level. We choose $u_m, \,v_m$ real and require $u_m \rightarrow 1$ as 
$m \rightarrow \infty$.
The coefficients $u_m,v_m$ are determined by 
numerically iterating the Hartree-Fock equations until a 
self-consistent solution is found.
Obviously, $|\Psi \rangle$ has electric charge $-Z|e|$ relative to the 
polarized groundstate 
$|\Psi_0 \rangle= \prod _{m=0}^{\infty} c^{\dagger}_{m\uparrow}|0\rangle$. 
Skyrmions with charge $Z<0$ are often referred to as antiskyrmions.
In the actual numerical calculation we have  a finite
number of particles, $m \leq N$, and impose $u_{N}=1$ (or $u_{N}=u_{N-1}=1$ see
below).

\noindent
{\bf Results:}
Our results are shown in Fig. 1-3. In Fig.~1 we give the energies to create
one $Z=2$ skyrmion and to create {\it two} $Z=1$ skyrmions, both at fixed particle
number and magnetic field, as functions of $\tilde g$. 
These energies are equal for the skyrmion ($Z>0$) and the antiskyrmion ($-Z<0$). 
This means that we have shifted the Hartree-Fock result for the state (\ref{hfstate})
by a constant.  For large $\tilde g$, the $Z=2$ skyrmion 
has higher energy than two $Z=1$ skyrmions. In this region the skyrmions are small, 
much smaller than the system size, and the Hartree-Fock calculation is reliable. 
\begin{figure*}[htbp]
  \begin{center}
  \leavevmode
    \epsfxsize = 8.6cm
    \epsfbox{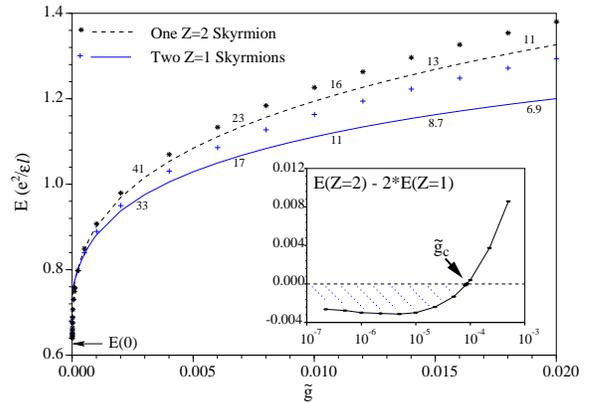}
  \end{center}
  \caption{Energies for one $Z=2$ skyrmion and for {\it two} $Z=1$ 
           skyrmions.
           The lines are Hartree-Fock results for $N=1000$ and the points are 
           sigma model results.
           The spin of one $Z=2$ and two $Z=1$ skyrmions is indicated along the lines. The 
           inset shows that the energies cross at  $\tilde g_c =9 \times 10^{-5}$ 
           (sigma model results only). $E(0)$ is the $\tilde g =0$ energy.} 

\label{energies}
\end{figure*}
The sigma model is to be trusted only for small $\tilde g$ when the skyrmions are large. 
It nevertheless gives energies that are close to, but slightly larger than, the 
Hartree-Fock energies for the range of  $\tilde g$ shown.  As $\tilde g$ decreases,
the difference in energy between one $Z=2$ and two $Z=1$ skyrmions decreases.
We see that, except possibly for very small $\tilde g$, the $Z=2$ skyrmion has 
larger energy. As noted earlier, in the limit $\tilde g \rightarrow 0$ the skyrmions 
become infinitely large and only the sigma model term in (\ref{effective}) contributes 
to the energy. Thus both energies must eventually approach the pure sigma model result 
$E(0)=8\pi \rho^s=\sqrt{\pi/8} \, e^2/\epsilon \ell$  as  $\tilde g \rightarrow 0$. 
To investigate the region of small $\tilde g$, we proceed in two different ways. 

The numerical sigma model method can be used for small 
$\tilde g$. The inset in Fig. 1 
shows the difference in energy as a function of $\tilde g$ obtained using the 
sigma model. This predicts a crossing at $\tilde g_c =
(9  \pm 1)\times 10^{-5}$. At this point the energy is 
$E_{\tilde g_c}=0.751 e^2/\epsilon \ell$  and there is a 10\% change in the spin 
per charge ($S/Z$) of the lowest energy excitation: $(S/Z)_{Z=1} =221$ and 
 $(S/Z)_{Z=2} =243$. The size of the excitation also changes: The charge radii 
(per unit charge) are $r_{Z=1}=21 \ell$ and $r_{Z=2}=31 \ell$. 
We can also see that when $\tilde g$ decreases below 
$\tilde g_c$,  the difference in energy first increases but then eventually starts to 
decrease as it must, since it should vanish at $\tilde g =0$.

Using Hartree-Fock for smaller $\tilde g$ than shown in Fig. 1, we find 
apparent crossing points, $\tilde g_c(N)$. However,  the $\tilde g_c(N)$
are so small that the skyrmions are large and there are large finite size 
effects: $\tilde g_c$
depends on $N$ and on whether one imposes as boundary condition 
that one or several of the spins are up at the edge of the system.  It is 
difficult to increase $N$ very much beyond $N=1000$ since the 
convergence is very slow for the small values of $\tilde g$ we are interested 
in. We conclude that 
within Hartree-Fock we are not able to {\it directly} probe a region of 
$\tilde g$ where an $N$-independent crossing might take place. 
However, we can reach smaller $\tilde g$ by finite size scaling.
In Fig. 2 we plot the crossing point, $\tilde g_c (N)$,
as a function of $1/N$ for three different sets of boundary conditions (as 
indicated in the figure) \cite{ask}. To each set we fit a quadratic polynomial in $1/N$ 
and read off $\tilde g_c \equiv \tilde g_c(\infty) =
(8.9 \pm 1.4)\times 10^{-5}$. The error is the 
standard deviation in 
the three extrapolated values. The errors in the fit 
for each curve are much smaller. Also, adding or subtracting a point at small 
$N$ does not affect the results. That the three 
different sets, which give very different $\tilde g_c (N)$, predict virtually 
the same nonzero $\tilde g_c$ indicates strongly that there is a crossing.
This value of $\tilde g_c$ agrees with the one obtained in the effective
theory. Thus we conclude that the doubly charged skyrmions have lower
energy for $\tilde g < \tilde g_c = 9  \times 10^{-5}$.

It is straightforward to calculate not only the energy but 
also the total spin as well as spin density and charge density. The total spin, 
which is indicated along the line in Fig. 1,   
is  much larger for the doubly charged skyrmion than for the charge one skyrmion. 
In Fig. 3 we give 
examples of densities for the doubly charged skyrmion and compare to {\it one} 
charge one skyrmion. Note that the charge density has its maximum a finite 
distance away from the origin for the doubly charged skyrmion (as the scale 
invariant skyrmion has).  

\begin{figure*}[htbp]
  \begin{center}
  \leavevmode
    \epsfxsize = 8.6cm
    \epsfbox{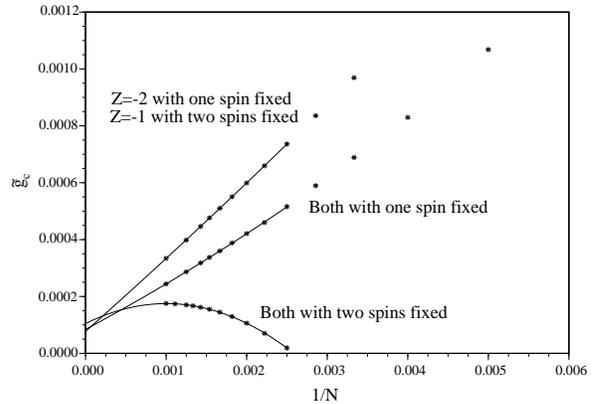}
  \end{center}
  \caption{Finite size scaling of  $\tilde g_c(N)$ using Hartree-Fock results 
           for antiskyrmions with  various boundary conditions (one or two spins up at the edge, 
           as indicated in the 
           figure). $\tilde g_c (N)$ is 
           where the energy of one $Z=-2$ skyrmion is equal to the energy of 
           two $Z=-1$ skyrmions in a Hartree-Fock calculation with $N$ particles. Note 
           that  $\tilde g_c (N)$ depends strongly on $N$ and on the boundary condition. 
           Fitting to  quadratic polynomials in $1/N$ gives
           $\tilde g_c \equiv \tilde g_c(\infty) = 8.9 \times 10^{-5}$. }
  \label{scaling}
\end{figure*}

\begin{figure*}[htbp]
  \begin{center}
  \leavevmode
    \epsfxsize = 8.6cm
    \epsfbox{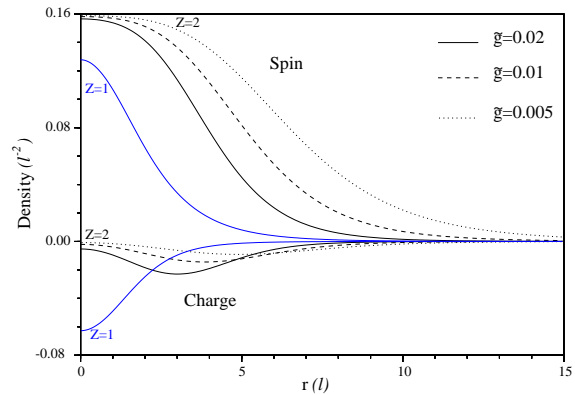}
  \end{center}
  \caption{Charge  and spin densities at $\tilde g = 0.02,\,0.01,\,0.005$ for 
           one $Z=2$ skyrmion (using Hartree-Fock). 
           Densities for one $Z=1$ skyrmion at $\tilde g = 0.02$ is
           included for comparison. }
\label{profiles}
\end{figure*}

The numerical sigma model gives $\lim _{\tilde g \rightarrow 0} 
\lbrack E-\sqrt{\pi/8}e^2/\epsilon \ell\rbrack /\tilde g ^{1/3}=
2.86 e^2/\epsilon \ell$ for $Z=2$ in agreement with (\ref{charge234}). Thus
there is no logarithm in the $Z=2$ energy and hence no reason to expect that
$Z>2$ skyrmions would ever have the lowest energy per charge. 
We have also checked that $Z=3$ skyrmions have higher energy 
(per charge) than $Z=1,2$ skyrmions for  $\tilde g > 10^{-7}$.

\noindent
{\bf Discussion and Experimental Implications:} Our analysis has
established that two skyrmions are unstable to pair-binding at very
small Zeeman energies. At distances much larger than their size,
the inter-skyrmion potential must be repulsive as it will be dominated
by their Coulomb interaction. It follows that the attraction that we
have found sets in only when the skyrmions overlap and their identities
begin to merge. In computing the skyrmion energies, we have ignored the
effects of Landau level mixing and of the softening of the Coulomb
interaction at small distances by the finite extent of the subband 
wavefunctions transverse to the plane of the electron gas. It is likely
that neither of these makes a huge difference to the energetics. In the
sigma model description, the dominant effect of Landau level mixing 
is a renormalization of the spin stiffness which does not affect the
inter-skyrmion energetics. The softening of the Coulomb interactions
is significant at distances which are $O(\ell)$ and for much bigger
skyrmions (such as in the region of interest where they exceed $20 \ell$)
most of the self-interaction is unsoftened. (We also note that the
long wavelength arguments regarding the peculiarities of the $Z=1$ solution
that underlie the pairing depend only on the form of the sigma model action; 
they will apply {\it mutatis mutandis} to other ferromagnetic fillings as well.)

This is encouraging in that it suggests that it might be possible to 
see the binding of the skyrmions as a sharp decrease in the spin
polarization with decreasing $\tilde g$ in the vicinity of $\nu=1$, where
the ground state will exhibit a dilute density of skyrmions. Unfortunately,
this is unlikely on two grounds. First, it would require an extremely fine 
tuning of $\tilde g$ which seems hard to accomplish. Second, there is the
more fundamental problem that when the skyrmions get to be very big, they
become sensitive to the details of the disorder potential which will tend
to limit their size. Effectively, we expect the disorder to provide
a lower limit to $\tilde g$ and while we have not computed this for realistic
disorder realizations, we are not greatly encouraged by the data in \cite{eaves}
which found a maximum skyrmion size, even for $\tilde g$ reduced under
pressure, considerably smaller than the critical one calculated in this paper.

\noindent{\bf Note Added:}
While finishing this work we became aware of a recent paper by Nazarov 
and Khaetskii \cite{nazarov} that studies higher topological charge skyrmions
in the effective sigma model in the spirit of our variational calculations. 
These authors also argue that the interaction between two charge one skyrmions 
is attractive at intermediate and short  distances and that the binding 
distance is zero, at which they form a charge two skyrmion. However, their quoted
results are independent of $\tilde g$, in contradiction with ours which hold
only at small $\tilde g$, and they appear not to have noticed the existence or the
extreme smallness of $\tilde g_c$.

\acknowledgements
We are grateful to S. A. Kivelson, R. Rajaraman and K. Yang  for useful
discussions. This work was supported in part by NSF grant No. DMR-9632690
and the A. P. Sloan Foundation (SLS) and the Swedish Natural Science Research 
Council (AK).


\begin{references}
\bibitem{meaculpa} Regrettably, the formula for the energy 
in Ref.~\cite{sondhi1} was  in
error by a factor of $1/2$  which we have corrected here.
\bibitem{ask} The data in  Fig. 2 are for antiskyrmions 
($Z<0$). The reason for this is technical: While the skyrmions and 
antiskyrmions have identical energies, it so happens that for very small 
$\tilde g$ the numerical errors in our calculations are much larger for the 
skyrmions.
\end{references}
\end{document}